\documentclass[aps,prb,twocolumn,superscriptaddress,showpacs]{revtex4-1}

\usepackage{graphicx}
\usepackage{calc}
\usepackage{bm}
\usepackage{color}
\usepackage{mathtools}

\begin{document}

\title{Magnetization symmetry for  MnTe  altermagnetic candidate}

\author{N.N. Orlova}
\author{V.D. Esin}
\author{A.V.~Timonina}
\author{N.N.~Kolesnikov}
\author{E.V.~Deviatov}
\affiliation{Institute of Solid State Physics of the Russian Academy of Sciences, Chernogolovka, Moscow District, 2 Academician Ossipyan str., 142432 Russia}

\date{\today}

\begin{abstract}
We experimentally investigate the magnetization angle dependence  $M(\alpha)$ for single crystals of MnTe altermagnetic candidate. In high magnetic fields, experimental $M(\alpha)$ curves mostly reflect standard antiferromagnetic spin-flop processes, which are allowed below the N\'eel vector reorientation field.  In low magnetic fields and at low temperatures,  spontaneous magnetization appears as a sharp $M(T)$ magnetization jump around 81~K. In this regime, $M(\alpha)$ dependence is quite unusual: the easy magnetization axis is  $\pi/2$ rotated either by increasing the field above 1~kOe or the temperature above 81~K. The observed behavior cannot be expected for antiferromagnetics, e.g. it differs strongly  from the well known weak ferromagnetism. Thus, it requires to take into account the formation of the altermagnetic ground state for MnTe altermagnetic candidate.  Despite MnTe is expected to have g-wave order parameter, $M(\alpha)$ magnetization symmetry  confirms the prevailing population of one from three easy axes, as it has been shown previously by temperature-dependent angle-resolved photo-emission  spectroscopy.
\end{abstract}

\maketitle

\section{Introduction}

Recently, a new class of altermagnetic materials has been added to  usual  ferro- and antiferro- magnetic classes~\cite{alter_common,alter_mazin}. Normally, ferromagnetic and antiferromagnetic materials  belong to the nonrelativistic groups of magnetic symmetry, i.e. to the case of weak spin-orbit coupling. In contrast, topological materials are always characterized by strong spin-orbit interaction~\cite{Volkov-Pankratov,MZHasan,Armitage}, and, therefore, by spin-momentum locking~\cite{sm-valley-locking}. For example,  spin is rotating along the Fermi-arcs in  Weyl semimetals~\cite{Fermi arc-SOC}, while the drumhead surface states lead to the spin textures of the skirmion type in topological nodal-line semimetals~\cite{nodal-line}. 

In altermagnets, the concept of spin-momentum locking~\cite{sm-valley-locking} was extended to the  non-relativistic groups of magnetic symmetry~\cite{alter_common,alter_mazin}. As a result, the small net magnetization is accompanied by alternating spin splitting in the k-space~\cite{alter_common,alter_josephson}. For example, for the d-wave altermagnet~\cite{AHE_RuO2} the up-polarized subband can be obtained by   $\pi/2$ rotation of the down-polarized one in the k-space~\cite{alter_supercond_notes,alter_normal_junction}.  There are also predictions  
 on the g- or even  i-wave order parameters~\cite{alter_common}, despite experimental verification is still an open question. For example, it was even proposed to address the k-space spin polarization by interfacing an altermagnet with the surface of a topological insulator~\cite{AHE_k_topol}.

Anomalous Hall effect~\cite{Armitage} was theoretically predicted~\cite{alter_original} and experimentally demonstrated~\cite{AHE_RuO2,AHE_MnTe1,AHE_MnTe2,AHE_Mn5Si3}  for altermagnets.  In contrast to RuO$_2$ altermagnetic candidate~\cite{AHE_RuO2}, the measurements in MnTe and Mn$_5$Si$_3$ show hysteresis and spontaneous AHE signals at zero magnetic field~\cite{AHE_MnTe1,AHE_MnTe2,AHE_Mn5Si3}. It is a common  agreement, that AHE still requires finite  net magnetization~\cite{satoru} and spin-orbit coupling also in altermagnetic materials~\cite{AHE_MnTe1,AHE_MnTe2,spin_ferro_soc}. 
The AHE transverse current is usually assumed to be perpendicular to  magnetization, therefore,  AHE and the ferromagnetic spin moment share the same symmetry. However, it is not generally so in materials with weak net magnetization~\cite{spin_ferro_soc,AHE_angle}. For example, the AHE signal does not correlate with the angle dependence of the weak saturation magnetization for Mn$_5$Si$_3$~\cite{AHE_Mn5Si3}.

\begin{figure}
\includegraphics[width=1\columnwidth]{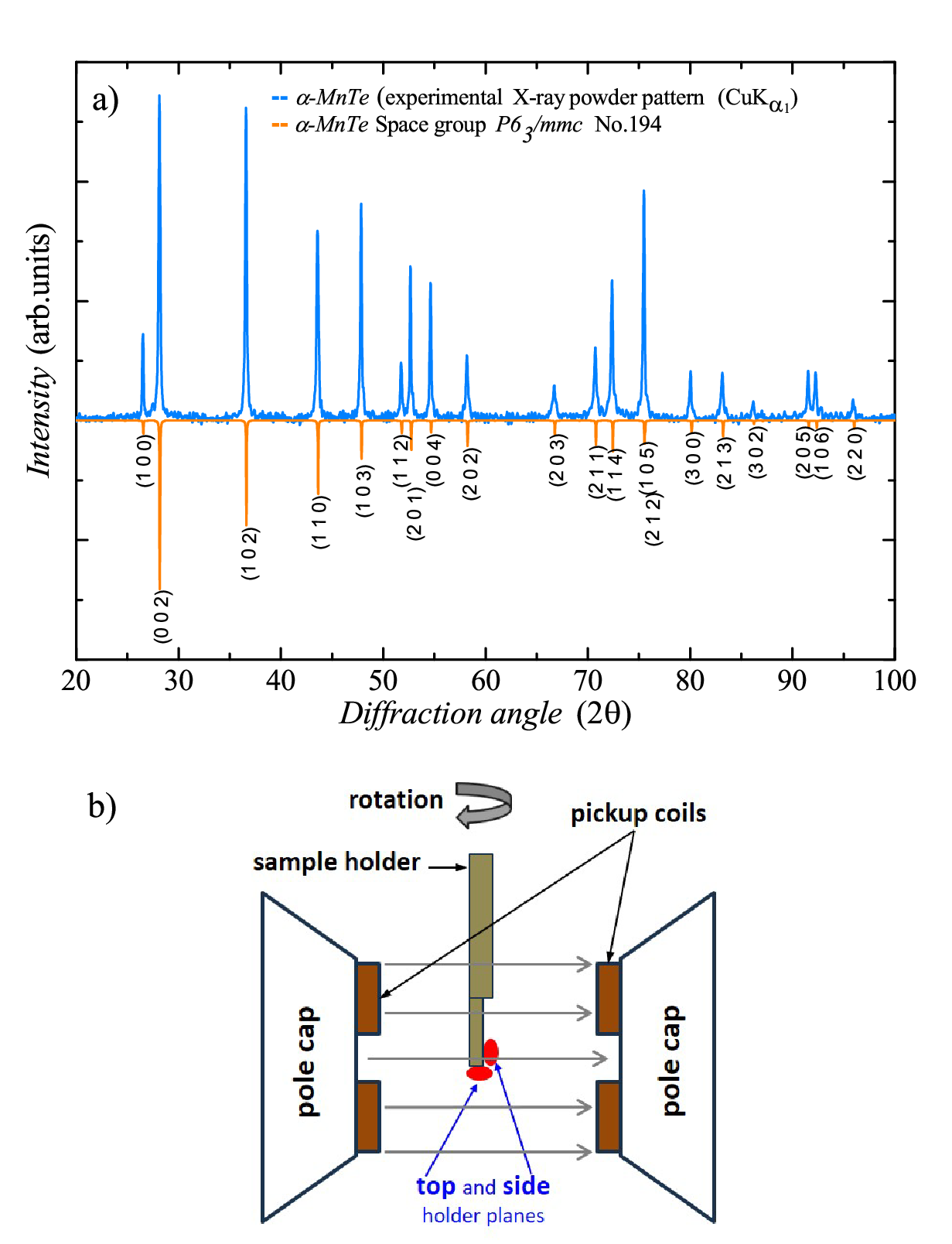}
\caption{(Color online) (a)  The X-ray powder diffraction  pattern (Cu K$_{\alpha1}$ radiation), which is obtained for the crushed MnTe single crystal. The single-phase  $\alpha$-MnTe is confirmed with the space group $P6_3 /mmc$ No. 194, so the results below cannot appear from the incorrect stoichiometry or oxides. 
(b) Schematic diagram of the experimental setup, with mutual orientation of the magnetic field, sample holder rotation axis and the magnetometer detector coils. The  coils are fixed to the magnet pole caps, so $M(\alpha)$ angle-dependent magnetization does not change a sign in constant magnetic field.  The sample can be mounted to the side or the top sample holder planes, to variate MnTe sample orientation in respect to the magnetic field and the rotation axis.  
  }
\label{sample}
\end{figure}

On the other hand, the order parameter is defined by the N\'eel vector symmetry in altermagnets~\cite{alter_common}, so the expected d-, g- or i-wave symmetry could be observable in magnetization measurements. This investigation can be conveniently performed for MnTe~\cite{MnTe_Mazin}, which 
is  characterized by accessible (2--3~T) magnetic field range~\cite{AHE_MnTe1,AHE_MnTe2} in contrast to RuO$_2$~\cite{AHE_RuO2} altermagnetic candidate. 

MnTe  is an intrinsic room-temperature magnetic semiconductor~\cite{AHE_MnTe1}.  The magnetic moments on Mn have a parallel alignment within the $c$ planes and an antiparallel alignment between the planes, so the two magnetic sublattices are connected by a sixfold screw axis along [0001]. 
Despite the long history of investigations~\cite{MnTe1,MnTe2,MnTe3,MnTe4,MnTe5,MnTe6}, the experimental data on the   magnetization angle dependence are quite controversial for MnTe. For example,  MnTe is of hexagonal NiAs structure, but $\pi/3$ periodicity has only been demonstrated  in high magnetic fields (much above 2~T) by torque technique~\cite{MnTe3} and by magnetoresistance measurements~\cite{MnTe_film_6phi}, while magnetization was only investigated for two perpendicular directions~\cite{MnTe_film_6phi}. Also, MnTe shows spontaneous ferromagnetic-like magnetization at low temperatures~\cite{orlova_mnte}, which is quite  unusual for collinear antiferromagnetics.  

As an altermagnetic candidate, MnTe is expected to have g-wave order parameter~\cite{alter_common,g_wave_MnTe}, although  it is still  debatable in recent publications~\cite{zyuzin}.  Thus, to conclude on the symmetry of the altermagnetic order parameter, it is reasonable to investigate the MnTe magnetization angle dependence in a wide field range. 

Here, we experimentally investigate the magnetization angle dependence  $M(\alpha)$ for single crystals of MnTe altermagnetic candidate. In high magnetic fields,  experimental $M(\alpha)$ curves mostly reflect standard antiferromagnetic spin-flop processes, which are allowed below the N\'eel vector reorientation field.  In low magnetic fields and at low temperatures,  spontaneous magnetization appears as a sharp $M(T)$ magnetization jump around 81~K. In this regime, $M(\alpha)$ dependence is quite unusual: the easy magnetization axis is  $\pi/2$ rotated either by increasing the field above 1~kOe or the temperature above 81~K. 

\section{Samples and technique}

To investigate magnetization anisotropy for $\alpha$--MnTe, it is preferable to use small MnTe single crystal samples rather than thin films. In the latter case, the results may be seriously affected by the geometrical factors and by the admixture of the substrate magnetic response, especially in low magnetic fields~\cite{MnTe_film_6phi}. For this reason, we investigate  small (2.89~mg and 5.73~mg mass) mechanically cleaved single crystal MnTe flakes.  Universality of the MnTe magnetization behavior is confirmed by high reproducibility  of the results, which are obtained for MnTe flakes of different mass and shape.

$\alpha$-MnTe was synthesized by reaction of elements (99.99\% Mn  and 99.9999\% Te) in evacuated silica ampules slowly heated up to 1050--1070$^\circ$C. The obtained loads were melted in the graphite crucibles under 10 MPa argon pressure, then homogenized at 1200$^\circ$C for 1 hour. The crystals grown by gradient freezing method are groups of single crystal domains with volume up to 0.5--1.0~cm$^3$. The MnTe composition is verified by energy-dispersive X-ray spectroscopy. The powder X-ray diffraction analysis confirms single-phase $\alpha$-MnTe with the space group $P6_3 /mmc$ No. 194, see Fig.~\ref{sample} (a).  As usual, the powder has been obtained by crashing the MnTe single crystal.

Sample magnetization  is measured by Lake Shore Cryotronics 8604 VSM magnetometer, equipped with nitrogen flow cryostat.  The magnetization anisotropy $M(\alpha)$ is investigated by sample holder rotation in the external magnetic field, see Fig.~\ref{sample} (b). The magnetometer detector coils are fixed to the magnet pole caps, so $M(\alpha)$ angle-dependent magnetization does not change a sign in constant magnetic field~\cite{orlova_mnte}.  Also, the sample can be mounted to either the side or the top sample holder planes   in Fig.~\ref{sample} (b), to variate MnTe sample orientation in respect to the magnetic field and the rotation axis.    Since for MnTe there is no definite cleavage plane, the initial orientation of the MnTe single crystal should be obtained from experimental  $M(\alpha)$ curves. 

The sample is mounted to the sample holder by low temperature grease. It is verified~\cite{gete}, that without MnTe sample, the sample holder with corresponding amount of grease shows fully  isotropic and strictly linear  diamagnetic response, which can be estimated as below  10\% of the measured MnTe magnetization value. Thus, the experimental setup allows to obtain magnetization angle dependence with high resolution in a wide magnetic field range $\pm15$~kOe. Before any measurements, the sample is cooled down the minimal 78~K temperature in zero magnetic field.  Afterward, the sample  is magnetized at 15~kOe, to have the stable, well-reproducible initial sample state.

\section{Experimental results}

\begin{figure}
\includegraphics[width=\columnwidth]{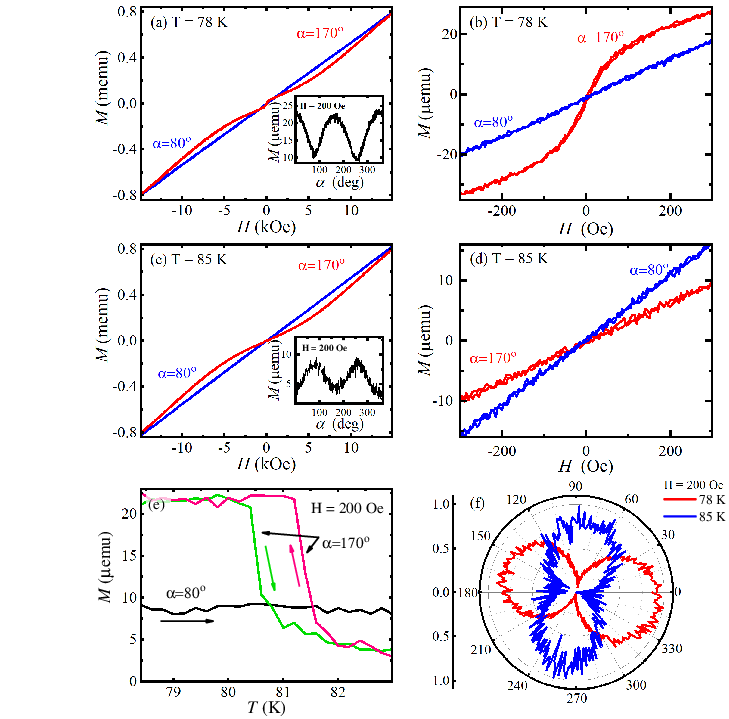}
\caption{(Color online)  $M(H)$ and $M(\alpha)$ magnetization dependences for the 2.89~mg MnTe sample. The sample is mounted to the side plane of the holder. (a) $M(H)$ magnetization loops within $\pm 15$~kOe field range at 78~K temperature. The curves are for M($\alpha$) maximum (the red curve, $\alpha$ = 170$^\circ$) and minimum (the blue curve, $\alpha$ = 80$^\circ$). The blue curve is strictly linear, while  there are nonlinear $M(H)$ branches with zero-field kink for the red one. Insert shows  $M(\alpha)$ angle dependence of magnetization, obtained for  200~Oe magnetic field at 78~K temperature. 
(b) The low-field $M(H)$ region at 78~K temperature, the curves are obtained with smaller magnetic field step. They confirm ferromagnetic-like  hysteresis at $\alpha$ = 170$^\circ$ (the red curve).
(c) and (d) Similarly obtained $M(H)$ and $M(\alpha)$ curves at 85~K temperature. $M(\alpha)$ is still $\pi$ periodic, see the inset, but the maxima and the minima positions are interchanged at 85~K in comparison with 78~K temperature.  The low-field $M(H)$ hysteresis disappears at 85~K. 
(e) The detailed $M(T)$ temperature dependences for two  orientations of 200~Oe magnetic field: the green and the pink curves are obtained at $\alpha$ = 170$^\circ$, for heating and cooling, respectively, while the  black  one is for $\alpha$ = 80$^\circ$. Spontaneous magnetization appears  as a sharp $M(T)$ jump for $\alpha$ = 170$^\circ$ only. 
(f)  The normalized  $\Delta M (\alpha)/M$  is shown as the circular diagram at 200~Oe magnetic field for two, 78~K and 85~K, temperatures. As the most important result, the easy axis is $\pi/2$ rotated when spontaneous  magnetization is destroyed above 81~K. 
 }
\label{fig2}
\end{figure} 

Fig.~\ref{fig2} shows $M(H)$ and $M(\alpha)$ magnetization dependences for the 2.89~mg MnTe sample. The sample is mounted to the side plane of the holder, as depicted in Fig.~\ref{sample} (b).

$M(\alpha)$ angle dependence of magnetization is demonstrated in the insert to Fig.~\ref{fig2} (a), it is obtained for  200~Oe magnetic field at 78~K temperature.  Two maxima ($\alpha$ = 170$^\circ$ and $\pi$-shifted) and two minima ($\alpha$ = 80$^\circ$ and $\pi$-shifted) indicate  $\pi$ periodicity of the experimental $M(\alpha)$ curve.

The main plot of Fig.~\ref{fig2} (a) shows $M(H)$ field-dependent sample magnetization loops within $\pm 15$~kOe field range at 78~K temperature.  $M(H)$ are measured by standard method of the magnetic field gradual sweeping between two opposite field values at fixed angle $\alpha$. The curves are for the M($\alpha$) maximum (the red curve, $\alpha$ = 170$^\circ$) and for the minimum (the blue curve, $\alpha$ = 80$^\circ$). In the latter case, $M(H)$ is strictly linear (the blue curve). In contrast, for $\alpha$ = 170$^\circ$, there are nonlinear $M(H)$ branches with pronounced zero-field kink (the red curve), similarly to Ref.~\cite{orlova_mnte}. 

The low-field $M(H)$ curves are shown in Fig.~\ref{fig2} (b), they are obtained with high accuracy for smaller magnetic field step. The curves confirm ferromagnetic-like  hysteresis at $\alpha$ = 170$^\circ$ (the red curve) and strictly linear $M(H)$ at $\alpha$ = 80$^\circ$ (the blue curve). Thus, at 78~K the low-field magnetization shows $\pi$-periodic easy-axis magnetic anisotropy.

Figs.~\ref{fig2} (c) and (d) show similarly obtained M(H) curves at 85~K temperature. $M(\alpha)$ is still $\pi$ periodic, see the inset to Fig.~\ref{fig2} (c), but the maxima and the minima positions are interchanged at 85~K in comparison with 78~K temperature (in the inset to Fig.~\ref{fig2} (a)). Also, $M(H)$ is strictly linear in low fields for both field orientations, so the low-field hysteresis disappears at 85~K, see  Fig.~\ref{fig2} (d). At higher fields, there are nonlinear $M(H)$ branches for $\alpha$ = 170$^\circ$ , which now corresponds to the M($\alpha$) minimum, while $M(H)$ is strictly linear for $\alpha$ = 80$^\circ$ (the blue curve). 

The detailed $M(T)$ temperature dependence is shown in Fig.~\ref{fig2} (e) for two above mentioned orientations of the 200~Oe magnetic field. The temperature is stabilized with  0.2~K step from 78.4~K to 83~K. The green and the pink curves are obtained at $\alpha$ = 170$^\circ$, for heating and cooling, respectively, the  black  one corresponds to $\alpha$ = 80$^\circ$. It can be seen from Fig.~\ref{fig2} (e), that ferromagnetic-like spontaneous magnetization appears below 80.5--81.5~K at $\alpha$ = 170$^\circ$ as a sharp $M(T)$ jump,  while magnetization is constant at  $\alpha$ = 80$^\circ$.   

$M(\alpha)$ results are summarized as the circular diagram in Fig.~\ref{fig2} (f). The normalized  $\Delta M/M$  is shown as a bar plot in polar coordinates at 200~Oe magnetic field for two, 78~K and 85~K, temperatures. Both the $\pi$-like periodicity and the easy-axis direction can be clearly seen from the diagram.  As the most important result, the easy axis is $\pi/2$ rotated when spontaneous ferromagnetic-like magnetization is destroyed above 81~K.

\begin{figure}
\includegraphics[width=\columnwidth]{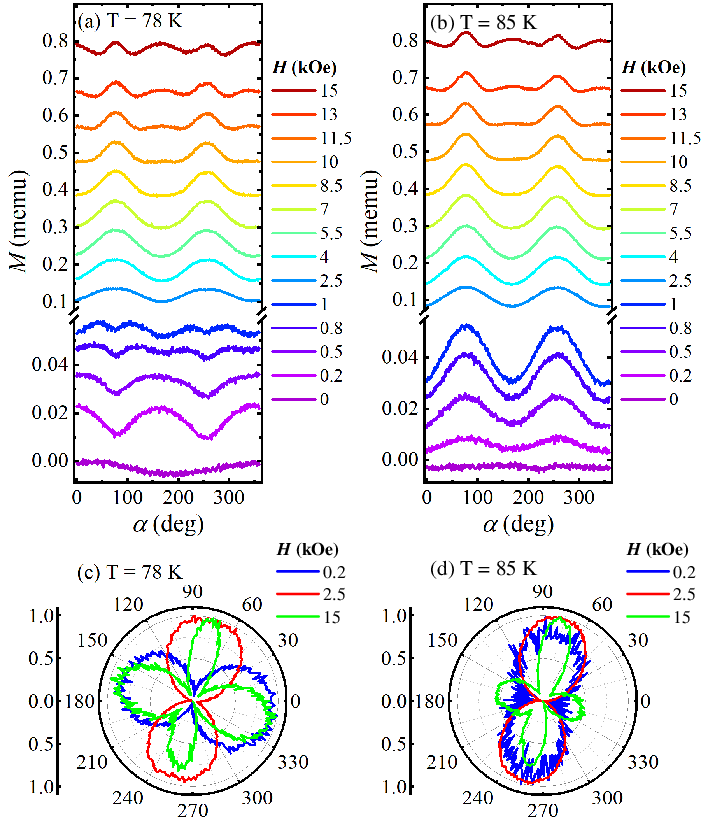}
\caption{(Color online) The detailed $M(\alpha)$ magnetization dependences for the 2.89~mg MnTe sample at  two, 78~K and 85~K, temperatures. 
In (a) and (b), the M($\alpha$)  curves are shown with  the 0.2--0.3~kOe step from 0 to the 1~kOe field value and with the 1.5--2~kOe step for the high,  1 to 15~kOe field range. From 2.5~kOe to 15~kOe, $M(\alpha)$ is of similar behavior for 78~K and at 85~K temperatures in (a) and (b), respectively. Below 1~kOe,  $M(\alpha)$ curves are qualitatively  different for 78~K and for 85~K temperatures: $M(\alpha)$ is always $\pi$-periodic, the maxima positions are stable in the full field range at 85~K. In contrast, $M(\alpha)$ is inverted at 78~K for the 0.2~kOe - 0.8~kOe magnetic fields.
(c) and (d) Circular diagrams at 78~K and at 85~K temperatures in (c) and (d), respectively, for the characteristic field values 0.2~kOe, 2.5~kOe and 15~kOe. The easy axis direction is $\pi/2$ rotated either by increasing the field above 1~kOe  in (c) or the temperature above 81~K in (d). 
 }
\label{fig3}
\end{figure}

Fig.~\ref{fig3} shows evolution of $M(\alpha)$ angular dependence of magnetization with magnetic field at two 78~K and 85~K temperatures, i.e. to both sides of the $M(T)$ jump at 81~K. The M($\alpha$)  curves are shown with  the 0.2--0.3~kOe step from 0 to the 1~kOe magnetic field and with the 1.5--2~kOe step for the higher fields. 

For high fields  from 2.5~kOe to 15~kOe, $M(\alpha)$ is of similar behavior for 78~K and at 85~K temperatures, see Figs.~\ref{fig3} (a) and (b), respectively. We observe $\pi$-periodicity of the curves  within the 2.5~kOe to 10~kOe  field range, while it is changed to the $\pi/2$-one with two additional maxima at higher fields. 

Below 1~kOe,  $M(\alpha)$ curves are qualitatively  different for 78~K and for 85~K temperatures in Figs.~\ref{fig3} (a) and (b). At  85~K , $M(\alpha)$ is always $\pi$-periodic, the maxima and minima positions are stable in the full field range, see Fig.~\ref{fig3} (b). In contrast, at 78~K temperature $M(\alpha)$ is just inverted for 0.2 --0.8~kOe magnetic fields in Fig.~\ref{fig3} (a), the maxima and the minima positions are interchanged in comparison with 85~K temperature.

At zero magnetic field, the remanence magnetization vanishes at 85~K temperature in Fig.~\ref{fig3} (b), while $M(H=0)$ clearly oscillates at 78~K  in Fig.~\ref{fig3} (a):  without the external field, constant sample magnetization produces $2\pi$-periodic oscillating signal in the magnetometer detector coils.
  
These $M(\alpha)$ data are summarized in the circular diagrams for the characteristic field values 0.2~kOe, 2.5~kOe and 15~kOe, see Figs.~\ref{fig3} (c) and (d). While increasing the magnetic field from 0.2~kOe to 2.5~kOe  in Fig.~\ref{fig3} (c),  the easy magnetization axis is rotating on the $\pi/2$ angle at 78~K. In contrast, the easy axis direction is stable at 85~K in Fig.~\ref{fig3} (d). At 15~kOe field, $M(\alpha)$ is of identical $\pi/2$-symmetry for both temperatures. Thus, the easy axis direction is $\pi/2$ rotated either by increasing the field above 1~kOe or the temperature above 81~K, see   Figs.~\ref{fig2} (f) and~\ref{fig3} (c-d).

\begin{figure}
\includegraphics[width=\columnwidth]{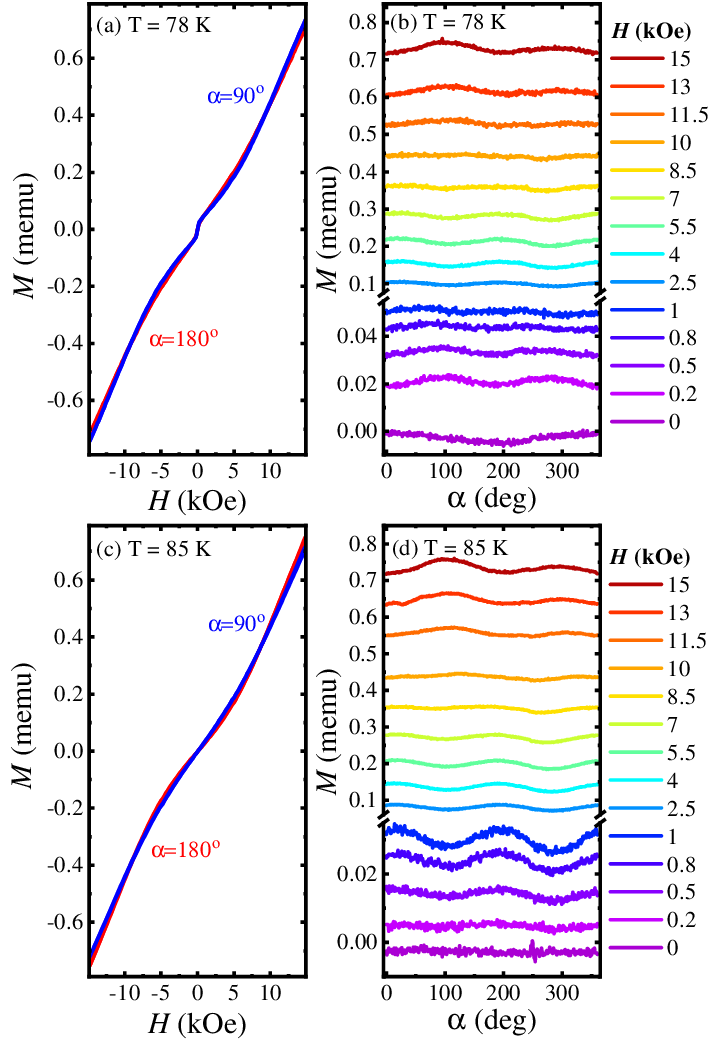}
\caption{(Color online) $M(H)$ and $M(\alpha)$ magnetization dependences for the 2.89~mg MnTe sample. The sample is fixed to the top holder plane by the same crystal surface, as depicted in Fig.~\ref{sample} (b). In general, the low-field behavior is equivalent for both sample orientations.
(a) $M(H)$ field-dependent magnetization within $\pm 15$~kOe field range at 78~K temperature. The curves are nonlinear for both $\alpha$ = 180$^\circ$ (the red curve) and $\alpha$ = 90$^\circ$ (the blue curve), so the magnetic field is always within the MnTe basal plane (see the text).  The low-field hysteresis is also present for both  angles. 
(b)  M($\alpha$)  curves at 78~K with  the 0.2--0.3~kOe step from 0 to the 1~kOe field  and with the 1.5--2~kOe step for higher  fields.
(c) Similarly obtained $M(H)$ curves at 85~K temperature for the same $\alpha$ values, the low-field hysteresis is suppressed. 
(d) M($\alpha$)  curves at 85~K. In low fields, $\pi$-periodic $M(\alpha)$ oscillations are inverted at 78~K in respect to 85~K, so the easy axis  is  $\pi/2$ rotated either by increasing the field to 2.5~kOe or the temperature to 85~K.
 }
\label{fig4}
\end{figure}  

To change the  MnTe sample orientation in respect to the rotation axis, and, therefore, to the magnetic field, the sample is fixed to the top holder plane by the same  crystal surface, as depicted in Fig.~\ref{sample} (b). The $M(H)$ and $M(\alpha)$ dependences are presented in Fig.~\ref{fig4}. From these data, external  magnetic field is always within  the N\'eel vector coplanar plane (the MnTe basal plane): the high-field  $M(H)$ branches are equally nonlinear for any angle $\alpha$.   MnTe is of hexagonal NiAs structure, the magnetic moments on Mn have a parallel alignment within the basal MnTe planes  and an antiparallel alignment between the planes. The nonlinear $M(H)$ branches are due to the antiferromagnetic domain configuration change~\cite{AFM_book}, which is only allowed for the field within the basal plane, as it can be seen in Fig.~\ref{fig4}. In contrast, $M(H)$ is strictly linear if the magnetic field is directed along the MnTe $c-$axis in Fig.~\ref{fig2} (a-d).  The overall $M(\alpha)$ oscillation amplitude is an order of magnitude smaller, there is no $\pi/2$ $M(\alpha)$ periodicity in high magnetic fields in Fig.~\ref{fig4} (b) and (d).

In general, the low-field behavior is equivalent for both sample orientations: there is no remanence magnetization at 85~K in Fig.~\ref{fig4} (d), while $M(H=0)$ oscillates at 78~K in Fig.~\ref{fig4} (b). The low-field hysteresis is also present for any rotation angle in  Fig.~\ref{fig4} (a), it still can be suppressed by temperature in   Fig.~\ref{fig4} (c). The  $\pi$-periodic $M(\alpha)$ oscillations are inverted at 78~K in comparison with 85~K, see Fig.~\ref{fig4} (b) and (d).

\begin{figure}
\includegraphics[width=\columnwidth]{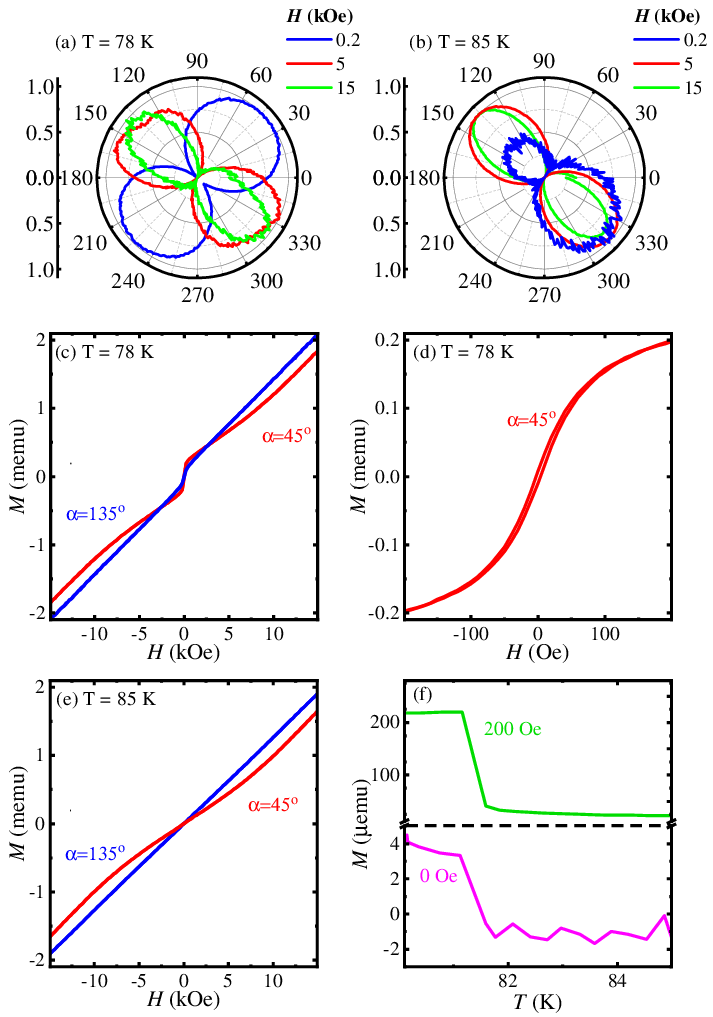}
\caption{(Color online) Similar $M(H)$ and $M(\alpha)$ magnetization behavior for the 5.73~mg  MnTe sample.
In (a) and (b),  $M(\alpha)$ dependences are summarized as circular diagrams for 78~K and 85~K temperatures, respectively.  For the low-temperature low-field ferromagnetic-like state, the easy axis direction is  $\pi/2$ rotated either by increasing the field above 1~kOe or the temperature above  81~K. 
(c) $M(H)$ curves with zero-field kink at  78~K temperature for two different angles $\alpha = 135^\circ$ (the blue curve) and  $\alpha = 45^\circ$ (the red curve). 
(d) The low-field $M(H)$ hysteresis at  78~K temperature for $\alpha = 45^\circ$. 
(e) The hysteresis disappears at 85~K, the high-field $M(H)$ branches are nonlinear for $\alpha = 45^\circ$ (the red curve), while $M(H)$ is nearly linear for  $\alpha = 135^\circ$ (the blue curve), so the field is parallel to the MnTe c-axis in the latter case.  
(f) Temperature transition as sharp $M(T)$ drop around 81~K. This drop can be observed both  for 200~Oe magnetic field (the green curve,  $\alpha = 45^\circ$), and for the remanence magnetization in zero external field (the magenta one).  
 }
\label{fig5}
\end{figure}

The reported behavior is sample-independent, it can be reproduced for any MnTe single crystal sample. For example,   Fig.~\ref{fig5} shows $M(H)$ and $M(\alpha)$ magnetization dependences for the 5.73~mg  MnTe sample. The results are summarized by circular diagrams in Fig.~\ref{fig5} (a) and (b), for 78~K and 85~K temperatures, respectively. For the low-temperature, low-field ferromagnetic-like state, the easy axis  is  $\pi/2$ rotated either by increasing the field above 1~kOe or the temperature above 81~K. 

At low temperatures, we observe nonlinear $M(H)$ curves with kink at zero field in Fig.~\ref{fig5} (c). The latter reflects the low-field hysteresis, as depicted in Fig.~\ref{fig5} (d). These features are suppressed at 85~K temperature in Fig.~\ref{fig5} (e), the transition is shown in Fig.~\ref{fig5} (f) as sharp $M(T)$ drop around 81~K. This drop can be observed both for  the $M(T)$ curve at 200~Oe  field, and for the $M(H=0)$ remanence magnetization in zero external magnetic field.  Also, magnetization value is proportional to the sample mass in Figs.~\ref{fig2} and ~\ref{fig5}, so the described behavior is due to the bulk MnTe properties.

\section{Discussion} \label{disc}

As a result,  $M(\alpha)$ is quite unusual  at low temperatures and in low magnetic fields: below 81~K, spontaneous magnetization appears, which is accompanied by easy axis rotation over $\pi/2$ by increasing the field above 1~kOe or the temperature above 81~K, see  Figs.~\ref{fig2} (f),~\ref{fig3} (c) and (d), and ~\ref{fig5} (a) and (b).  

Let us start from the high-field behavior.  We argue, that while the nonlinear $M(H)$ branches are due to the antiferromagnetic domain configuration change, the observed $\pi$ or $\pi/2$ periodicity in high magnetic fields reflects usual antiferromagnetic spin-flop  below the N\'eel vector reorientation field~\cite{MnTe_film_6phi,AFM_book}. Spin-flop  is known as simultaneous canting of two magnetic sublattices due to the external magnetic field. In moderate fields, it is only allowed within the MnTe basal plane, which conserves the easy-axis anisotropy (i.e. $\pi$-periodic  $M(\alpha)$ for 1~kOe - 10~kOe fields in Fig.~\ref{fig3}). For higher fields, spin-flops are also allowed along the $c$-axis~\cite{AFM_book}, which leads to more complicated $\pi/2$ periodicity in Fig.~\ref{fig3} above 10~kOe. For this reason, we do not observe $\pi/2$ periodicity for the magnetic field within the MnTe basal plane in Fig.~\ref{fig4}. 

In low fields, we do not observe $\pi/3$ periodicity of magnetization for any sample orientation, despite it should be expected for the hexagonal MnTe structure. This discrepancy can not be connected with  any sample disadvantages, like incorrect MnTe stoichiometry, Mn oxides, defects, etc~\cite{impur1,impur2}. For our samples, the MnTe composition was verified by energy-dispersive X-ray spectroscopy and the powder X-ray diffraction analysis. The latter also confirms the space group $P6_3 /mmc$ No. 194, see Fig.~\ref{sample} (a), so one should be sure in $\pi/3$ symmetry for the MnTe basal plane.  Moreover, $\pi/3$ periodicity of magnetization has never been demonstrated for MnTe except two indirect high-field experiments~\cite{MnTe3,MnTe_film_6phi}, despite recent magnetoresistance measurements still show  $\pi$ periodicity at least below the N\'eel vector reorientation field~\cite{dash_MnTe_magres}.

As a result, the  difference between the $M(\alpha)$ magnetization symmetry and the MnTe crystal structure requires to take into account the electronic properties, i.e. formation of the altermagnetic ground state. 

MnTe belongs to a new class of altermagnetic materials~\cite{MnTe_Mazin,AHE_MnTe1,AHE_MnTe2}, the small net magnetization is accompanied by alternating spin-momentum locking in the k-space, so the unusual spin splitting is predicted~\cite{alter_common,alter_josephson}. For the altermagnetic candidate MnTe, it is accepted, that the principle origin of nonzero net magnetic moment~\cite{AHE_MnTe1,AHE_MnTe2,orlova_mnte}, and, therefore,  of weak remanence  magnetization~\cite{alter_ferro,spin_ferro_soc} is the spin-orbit coupling~\cite{satoru} in the valence orbitals~\cite{Dichroism}.  The effects of spin-orbit coupling in this material has been previously investigated by temperature-dependent angle-resolved photo-emission spectroscopy (ARPES) and by disordered local moment calculations~\cite{MnTe_SO}.

In contrast to  the expected g-wave MnTe altermagnetic state ,  our experiment demonstrate $\pi$  periodicity of magnetization in low magnetic fields, i.e. d-wave order parameter (see also theoretical discussion in Ref.~\cite{zyuzin}).  For the altermagnetic state in bulk MnTe samples,  the prevailing population of one from three easy axes has been shown by ARPES~\cite{MnTe_ARPES2}. A lower twofold symmetry has therefore been  established at energies near the top of the valence band for strong spin-orbit coupling, which confirms d-wave order parameter ($\pi$  periodicity of magnetization).  In recent Ref.~\cite{theor_new_MnTe} similar result has been obtained theoretically for MnTe. 

It seems to be important, that spin-orbit coupling is well-resolved only below 100~K in Ref.~~\cite{MnTe_SO}, which is consistent with appearance of spontaneous magnetization  in Figs.~\ref{fig2} (e) and~\ref{fig5} (f).  Thus, above the 81~K transition temperature or in higher magnetic fields, the easy-axis pinning to the crystal structure is recovered, which is  responsible for the easy axis $\pi/2$ rotation either by increasing the field above 1~kOe or the temperature above 80.5--81.5~K in  Figs.~\ref{fig2} (f),~\ref{fig3} (c) and (d). 

We wish to note, that  spontaneous magnetization of MnTe below 81~K is strongly different from the well known weak ferromagnetism~\cite{weak_ferro,weak_ferro1} of conventional antiferromagnetics:

(i) MnTe space group is different from the required for  weak ferromagnetism~\cite{weak_ferro1,symmetry_alter}.  

(ii) Weak ferromagtetism originates from the symmetry lowering~\cite{weak_ferro1} in high magnetic fields ( above 2~kOe range), it appears  as the small, 0.1~\%-10~\% $M(H)$ jump~\cite{weak_ferro}. In contrast, we observe remanence magnetization below 81~K without any external field.   In low 200~Oe field, $M(\alpha)$ modulation is about 50\% in Figs.~\ref{fig2}  and~\ref{fig3}.

(iii) Temperature transition as  sharp $M(T)$ jump is very unusual for standard ferromagnetic transitions.  It also excludes the impurity-induced ferromagnetism, in addition to the correct MnTe stoichiometry for our samples. 

Thus, despite the ground state symmetry is still debatable in MnTe~\cite{MnTe_ARPES2}, the easy axis rotation over $\pi/2$ in Figs.~\ref{fig2} (f),~\ref{fig3} (c) and (d) seems to be the crucial experimental argument for the MnTe altermagnetic ground state below 81~K.

\section{Conclusion}
As a conclusion, we experimentally investigate the magnetization angle dependence  $M(\alpha)$ for single crystals of MnTe altermagnetic candidate. In high magnetic fields, experimental $M(\alpha)$ curves mostly reflect standard antiferromagnetic spin-flop processes, which are allowed below the N\'eel vector reorientation field.  In low magnetic fields and at low temperatures,  spontaneous magnetization appears as a sharp $M(T)$ magnetization jump around 81~K. In this regime, $M(\alpha)$ dependence is quite unusual: the easy magnetization axis is  $\pi/2$ rotated either by increasing the field above 1~kOe or the temperature above 81~K. The observed behavior cannot be expected for antiferromagnetics, e.g. it differs strongly  from the well known weak ferromagnetism. Thus, it requires to take into account the formation of the altermagnetic ground state for MnTe altermagnetic candidate.  Despite MnTe is expected to have g-wave order parameter, $M(\alpha)$ magnetization symmetry  confirms the prevailing population of one from three easy axes, as it has been shown previously by temperature-dependent angle-resolved photo-emission  spectroscopy.

\acknowledgments

We wish to thank S.S~Khasanov for X-ray sample characterization and Vladimir Zyuzin for valuable discussions.  We gratefully acknowledge financial support  by the  Russian Science Foundation, project RSF-24-22-00060, https://rscf.ru/project/24-22-00060/

\end{document}